\documentclass[AMA,STIX1COL,doublespace]{WileyNJD-v2}

\usepackage{graphicx}
\usepackage{amsmath}

\usepackage{times}
\usepackage{bm}
\usepackage{enumerate}
\usepackage{xurl}

\usepackage{mathtools}
\usepackage{setspace}
\usepackage{array,booktabs}
\usepackage{tikz}
\usepackage{mathtools}
\usepackage{chngcntr}
\usepackage{algorithm}
\usepackage{algpseudocode}
\usepackage{multirow}
\usepackage{multicol}
\usepackage[font=normalsize]{subcaption}
\usepackage{amsmath}
\usepackage{footnote}
\usepackage{enumitem}
\usepackage{amsmath}
\usepackage{soul}

\articletype{RESEARCH ARTICLE}%

\received{29 January 2021}
\revised{14 April 2021}
\accepted{27 May 2021}

\begin{document}

\title{Synergy Area with FDR-controlled Evaluation (SAFE) to robustly assess safety profile in clinical trials}

\author{Tianyu Zhan$^{1,*}$, Yabing Mai$^{1}$, Yihua Gu$^{1}$, Thao Doan$^{2}$ and Xun Chen$^{1}$ }

\authormark{Zhan et al.}

\address{	$^{1}$Data and Statistical Sciences, AbbVie Inc., 1 Waukegan Road, North Chicago, IL 60064, U.S.A. \\
	$^{2}$Patient Safety, AbbVie Inc., 1 Waukegan Road, North Chicago, IL 60064, U.S.A.}

\corres{Tianyu Zhan, 1 Waukegan Rd, North Chicago, IL 60064, USA. \\ \email{tianyu.zhan.stats@gmail.com}}

%\presentaddress{This is sample for present address text this is sample for present address text}

\abstract[Summary]{
Safety assessment plays a fundamental role in developing a new drug via clinical trials for ethical considerations. Due to complexity, manual review is typically conducted on the totality of data to draw safety conclusions. There are some existing quantitative methods to facilitate or tailor further medical review, with a controlled error rate and integration of clinical knowledge. In addition to those two key aspects, we emphasize the importance of relying on substantial evidence to draw robust conclusions on safety. Motivated by these three important properties, we propose a two-layer Synergy Area with FDR-controlled Evaluation (SAFE) structural framework to robustly assess the safety profile in clinical trials. In the first layer of SAFE, we investigate each clinically meaningful Synergy Area (SA) based on compelling evidence. In the next layer, the false discovery rate (FDR) is controlled for potential findings across all SAs. Simulation studies show that SAFE properly controls error rates within and across SAs at the nominal level. We further apply the proposed approach to two case studies based on real data from the Historical Trial Data (HTD) Sharing Initiative of the DataCelerate platform. As compared to some direct methods, SAFE demonstrates an appealing feature of screening out extreme data and reaching solid safety conclusions. It can act as either a building block in another framework, or a platform to incorporate additional components. 
}


%\jnlcitation{\cname{%
%\author{Williams K.}, 
%\author{B. Hoskins}, 
%\author{R. Lee}, 
%\author{G. Masato}, and 
%\author{T. Woollings}} (\cyear{2016}), 
%\ctitle{A regime analysis of Atlantic winter jet variability applied to evaluate HadGEM3-GC2}, \cjournal{Q.J.R. Meteorol. Soc.}, \cvol{2017;00:1--6}.}

\maketitle

%\footnotetext{\textbf{Abbreviations:} ANA, anti-nuclear antibodies; APC, antigen-presenting cells; IRF, interferon regulatory factor}

\section{Introduction}
\label{sec:intro}

When investigating a new biological treatment in clinical trials, its safety is at least as important as efficacy.\cite{talbot2008efficacy} The evaluation of efficacy is relatively straightforward to translate to several logical or statistical problems, such as endpoint derivation, point estimation, hypothesis testing, and multiple comparisons. However, a comprehensive and accurate assessment of safety is much more complicated \citep{singh2012drug}, not just due to a relatively large number of adverse event (AE) terms. For a specific reported AE, it contains extensive additional information, such as severity, serious or not, causality to the study drug, action taken for the study drug, related laboratory values and narratives. \citep{simon2019principles} Therefore, a standard approach is to generate some relevant standard summaries (e.g., incidence rates, exposure-adjusted rates, listings, narratives), and then conduct a detailed clinical review on the totality of data.\citep{fdasafety}

Some previous work has been done to provide certain quantitative assistance to the heavy, tedious, and even subjective manual review. A natural stream is to directly apply multiplicity adjustment to control either family-wise error rate (FWER) or false discovery rate (FDR) based on results from all AEs or other categories.\citep{huque2010multiplicity, dmitrienko2013traditional, norton2016perspective, menyhart2025multiplicity} Clinical and other domain knowledge, such as biological relationships between AEs, is also critical to safety assessments. Some other methods take those clinical aspects into consideration \citep{berry2004accounting, dunson2008bayesian, berry2010bayesian, mehrotra2012flagging}, mainly using a Bayesian approach. 

In addition to those two important aspects of error rate control and domain knowledge, we emphasize another point: relying on substantial evidence to draw robust safety conclusions. Multiple clinically correlated findings are stronger than a single finding, even if it is extremely significant after multiplicity adjustment. This philosophy is well-accepted and commonly used in clinical practice. For example, the diagnosis of a heart attack is based on several symptoms and biomarkers.\citep{daubert2010utility} As a summary, a proper quantitative safety evaluation approach requires at least three important features: 1. a proper error rate control; 2. incorporation of clinical and domain knowledge; 3. robust conclusions based on substantial evidence. Most previous quantitative frameworks\citep{huque2010multiplicity, dmitrienko2013traditional, norton2016perspective, menyhart2025multiplicity, berry2004accounting, dunson2008bayesian, berry2010bayesian, mehrotra2012flagging, de2024statistical} focus on the first two properties, but not necessarily the third one. 

To accommodate those three critical aspects, especially the third one, we propose a Synergy Area with FDR-controlled Evaluation (SAFE) structural framework to robustly and automatically assess the safety profile in clinical trials. It naturally incorporates clinical and other domain knowledge to define a class of Synergy Areas (SAs), such as body regions. SAFE has an overall two-layer structure. In the first layer within a specific SA, we declare that this SA has a positive finding if at least two AE variables have signals. It can also be flexibly generalized to another setting, say, at least four elementary discoveries, to fit the purpose of another problem. This feature echoes the third aspect discussed above to draw robust conclusions based on substantial evidence. The naming of "Synergy Area (SA)" reflects this spirit of relying on at least two findings to draw a more robust conclusion. In the second layer of SAFE to provide some regulations, we control the false discovery rate (FDR), which is more appealing to exploratory research \citep{benjamini1995controlling} as in our context of safety assessment. Potential safety findings based on the SAFE framework trigger additional and extensive medical review to confirm conclusions, and can also motivate further in-depth and tailored investigations. Integrating two layers of SAFE and some existing multiplicity adjustment methods \citep{holm1979simple, benjamini1995controlling, benjamini2001control}, we propose an algorithm to properly control FDR across SAs, and the error rate within an SA. 

The rest of this manuscript is organized as follows. Section \ref{sec:safe} demonstrates the SAFE framework with the bottom-up approach. In Section \ref{sec:alter}, some alternative methods are discussed. Simulation studies are conducted in Section \ref{sec:sim} to evaluate error rate control of SAFE under varying settings. In Section \ref{sec:real}, we apply this method to two case studies based on real clinical trial data from the Historical Trial Data (HTD) Sharing Initiative of TransCelerate BioPharma. Additional discussion and future work are presented in Section \ref{sec:disc}.

\section{Synergy Area with FDR-controlled Evaluation (SAFE) Framework}
\label{sec:safe}

\subsection{A two-layer structure}

Suppose that there are $m$ Synergy Areas (SAs) of interest, which are denoted as $\text{SA}_1$, $\text{SA}_2$, ..., $\text{SA}_m$. The rationale behind this name will be discussed in Section \ref{sec:first}. Its specific choice is flexible, and depends on the clinical interpretation of the safety profile. For example, one may use the System Organ Class (SOC), which is the highest level of adverse event (AE) coding terms in MedDRA (Medical Dictionary for Regulatory Activities)\cite{meddra}, or the high-level group term in MedDRA. Some other candidate classifications are: the severity scale of AE, causality of AE to the study drug, corresponding action taken with the study drug, etc. Within each $\text{SA}_i$ ($i = 1, ..., m$), we have $n_i$ AE variables $V_{i, 1}$, $V_{i, 2}$, ..., $V_{i, n_i}$. 

\begin{figure}[h]
	\centering
	\includegraphics[width=16cm]{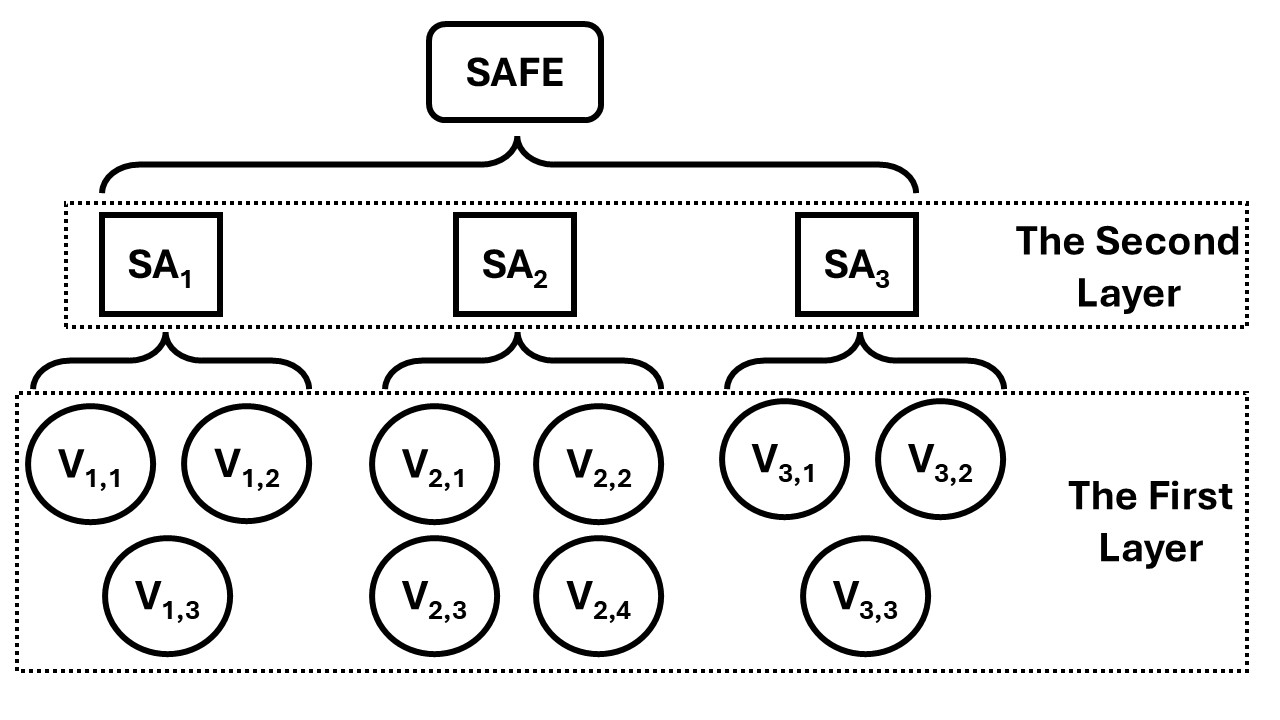}
	\caption{A graphical illustration of the two-layer Synergy Area with FDR-controlled Evaluation (SAFE) framework. There are three Synergy Areas (SAs) with $m=3$: $\text{SA}_1$, $\text{SA}_2$ and $\text{SA}_3$. Within $\text{SA}_1$, we have three AE variables with $n_1 = 3$: $V_{1, 1}$, $V_{1, 2}$ and $V_{1, 3}$. Correspondingly, $\text{SA}_2$ has $n_2 = 4$ AE variables, and $\text{SA}_3$ has $n_3 = 3$ AE variables. }
	\label{fig:diagram}
\end{figure}

Figure \ref{fig:diagram} provides a graphical diagram of this two-layer Synergy Area with FDR-controlled Evaluation (SAFE) framework to assess the safety profile. In this particular example, we have a total of $m = 3$ SAs in the top layer. At bottom, we have $n_1 = 3$, $n_2 = 4$, and $n_3 = 3$ AE variables within $\text{SA}_1$, $\text{SA}_2$ and $\text{SA}_3$, respectively. 

\subsection{The first layer within a Synergy Area (SA)}
\label{sec:first}

We use the bottom-up approach to introduce our proposed framework. To begin with, we discuss the first layer of AE variables within a specific SA. For illustration, we consider $\text{SA}_i$ with index $i$ and associated variables $V_{i, j}$ ($j = 1, ..., n_i$).

For the variable $V_{i, j}$, we denote $H_0^{(i, j)}$ as the null hypothesis with no abnormal safety signals, $H_1^{(i, j)}$ as the alternative hypothesis, and $p_{i, j}$ as the corresponding \textit{p}-value. The computation of $p_{i, j}$ depends on the specific context of a problem and the parameter of interest. For example, in a two-group randomized clinical trial with the AE incidence rate as the parameter to inform the safety profile, we can compute the $p$-value based on the two-group proportion test, as in Section \ref{sec:real}. 

As we move on to the Synergy Area $\text{SA}_i$, typically one formulates the null hypothesis as $K_0^{(i)}$ and the alternative hypothesis as $K_1^{(i)}$: 
\begin{align}
K_0^{(i)} & = H_0^{(i, 1)} \cap H_0^{(i, 2)} \cap ... \cap H_0^{(i, n_i)}, \label{equ:g0_ori} \\
K_1^{(i)} & = H_1^{(i, 1)} \cup H_1^{(i, 2)} \cup ... \cup H_1^{(i, n_i)}, \label{equ:g1_ori}
\end{align}
where the null hypothesis $K_0^{(i)}$ for $\text{SA}_i$ is the intersection of all elementary null hypotheses $H_0^{(i, j)}$'s, and the alternative hypothesis $K_1^{(i)}$ is the union of all elementary alternative hypotheses $H_1^{(i, j)}$'s. Hypothesis testing of $K_1^{(i)}$ versus $K_0^{(i)}$ is conducted based on raw \textit{p}-values $p_{i, j}$'s adjusted by multiple test procedures, such as the Holm procedure.\citep{holm1979simple} 

However, the alternative hypothesis $K_1^{(i)}$ in (\ref{equ:g1_ori}) is true, even if just one $H_1^{(i, j)}$ for a single AE variable is true within a given $\text{SA}_i$. To enhance robustness of decision making, we consider the following setups of the null hypothesis $G_0^{(i)}$ and the alternative hypothesis $G_1^{(i)}$: 
\begin{align}
G_0^{(i)} & = \overline{G}_1^{(i)}, \label{equ:g0_new} \\
G_1^{(i)} & = \underset{1 \leq j < k \leq n_i}{\cup} \left[ H_1^{(i, j)} \cap H_1^{(i, k)} \right], \label{equ:g1_new}
\end{align}
where $\overline{G}$ is the complement of $G$. For example, suppose that there are $n_i=3$ AEs within $\text{SA}_i$, then $G_1^{(i)}$ is $ \left[ H_1^{(i, 1)} \cap H_1^{(i, 2)} \right] \cup  \left[ H_1^{(i, 1)} \cap H_1^{(i, 3)} \right] \cup  \left[ H_1^{(i, 2)} \cap H_1^{(i, 3)} \right]$.

The alternative hypothesis $G_1^{(i)}$ in (\ref{equ:g1_new}) is valid if at least two elementary alternative hypotheses $H_1^{(i, j)}$ and $H_1^{(i, k)}$ are true. That is to say, this $\text{SA}_i$ is labeled with abnormal safety signals if at least two AE variables within this SA have findings. One example is the diagnosis of acute myocardial infarction (or sometimes referred to as heart attack), which is usually based on abnormal troponin as a cardiac biomarker and some additional conditions, such as chest pain and electrocardiogram (ECG) changes.\citep{daubert2010utility} Criteria for Hy's law of assessing liver injury include several quantitative conditions with alanine aminotransferase, aspartate aminotransferase and total bilirubin.\citep{re2015risk} From a clinical perspective, several clinically correlated symptoms are needed to confirm a safety conclusion. Therefore, we propose to focus on the alternative hypothesis $G_1^{(i)}$ in (\ref{equ:g1_new}) with at least two findings. Additionally, this rationale also motivates the naming of SA as ``Synergy Area''. This setup can be generalized to other settings, such as at least four findings within an SA. 

When it comes to hypothesis testing, we first compute adjusted $p$-values $\widetilde{p}_{i, j}$ based on raw $p$-values ${p}_{i, j}$ by the Holm procedure.\citep{holm1979simple} The protection of family-wise error rate (FWER) by the Holm procedure is free from the potential unknown dependence structure of those AE variables.\citep{holm1979simple} This feature is especially appealing to safety assessment, because some variables may be negatively correlated. If there is a strong knowledge that variables are under some positive dependency, then other more powerful multiplicity-adjustment methods can be applied, such as the Hommel procedure.\citep{hommel1988stagewise}

Then we define the decision function  $D_i(\widetilde{\alpha})$ for hypothesis testing in $\text{SA}_i$ as, 
\begin{equation}
\label{equ:dec}
D_i(\widetilde{\alpha}) = I \left( \widetilde{p}^{(2)}_{i} \leq \widetilde{\alpha} \right),
\end{equation}
which takes a value of $1$ if there is a safety concern in this Synergy Area (SA) $i$, and is equal to $0$ otherwise. The significance level $\widetilde{\alpha}$ is a constant between $0$ and $1$. The value $\widetilde{p}^{(k)}_{i}$ is the $k$th element of the sorted adjusted $p$-values $\left(\widetilde{p}_{i, 1}, \widetilde{p}_{i, 2}, ..., \widetilde{p}_{i, n_i}\right)$ in ascending order. Specifically, $\widetilde{p}^{(2)}_{i}$ is the second smallest adjusted $p$-value. The component of $\widetilde{p}^{(2)}_{i} \leq \widetilde{\alpha}$ in (\ref{equ:dec}) is equivalent to the event of having at least two adjusted $p$-values smaller than or equal to the nominal level $\widetilde{\alpha}$. This decision function is aligned with the alternative hypothesis $G_1^{(i)}$ in (\ref{equ:g1_new}). 

On the error rate control, this $D_i(\widetilde{\alpha})$ offers a proper control of FWER at $\widetilde{\alpha}$ with adjusted $p$-values from the Holm procedure when testing $K_1^{(i)}$ in (\ref{equ:g1_ori}) versus $K_0^{(i)}$ in (\ref{equ:g0_ori}). There is no further adjustment needed, because $D_i(\widetilde{\alpha})$ requires a more stringent condition with the second smallest adjusted $p$-value $\widetilde{p}^{(2)}_{i}$ than the traditional setting of the Holm procedure with the smallest adjusted $p$-value $\widetilde{p}^{(1)}_{i}$. On a more stringent alternative hypothesis $G_1^{(i)}$ in (\ref{equ:g1_new}), it has been shown that the same decision function $D_i(\widetilde{\alpha})$ also properly controls the error rate at $\widetilde{\alpha}$ without additional adjustment.\cite{berger1982multiparameter} Numerical results in Section \ref{sec:sim} further confirm such a control, and demonstrate scenarios where the error rate is close to the nominal level $\widetilde{\alpha}$. Those findings are consistent with some theoretical results.\cite{berger1982multiparameter}

Of note, this setup of (\ref{equ:g0_new}) and (\ref{equ:g1_new}) is closely related to the partial conjunction test\citep{benjamini2008screening}. Under a general dependency structure, their Bonferroni-type p-value combining procedure\citep{benjamini2008screening} is equivalent to our approach in (\ref{equ:dec}). In this work, we focus on such a robust strategy with proper error rate control under arbitrary dependency. This framework can be extended to more powerful testing procedures with additional conditions assumed, e.g., the shifted Simes p-value under three conditions.\citep{benjamini2008screening} 

As a summary of this section, for each of the $\text{SA}_i$, we have constructed a level-$\widetilde{\alpha}$ test $D_i(\widetilde{\alpha})$ in (\ref{equ:dec}) to test $G_1^{(i)}$ in (\ref{equ:g1_new}) versus $G_0^{(i)}$ in (\ref{equ:g0_new}). 

\subsection{The second layer across SAs}
\label{sec:second}

In the second layer, our objective is to identify potential safety signals from all $m$ SAs with a controlled false discovery rate (FDR)\cite{benjamini1995controlling} at a nominal level of $\alpha$, which is a constant between $0$ and $1$. We first provide a brief review of FDR based on some previous works \citep{benjamini1995controlling} in the context of this article. FDR is defined as,
\begin{equation}
\label{equ:FDR}
\text{FDR} = E(Q) = E\left(S/R \right),
\end{equation}
where $Q = S/R$, $S$ is the number of wrongly identified SAs with safety findings, and $R$ is the total number of SAs with identified safety signals. When $R=0$ with no positive findings, we set $Q=0$ because there is no false positive error. FDR in (\ref{equ:FDR}) is interpreted as the expected proportion of falsely detected SAs among all discoveries. When all SAs have no safety signals, FDR is equivalent to FWER, which is the probability of falsely identifying at least one SA. 

Suppose that all SAs have associated \textit{p}-values $\breve{p}_i$ ($i = 1, ..., m$), then the Benjamini-Hochberg (BH) procedure is able to control FDR at $\alpha$ when all $\breve{p}_i$'s are independent from each other.\citep{benjamini1995controlling} To conduct the hypothesis testing, we compute $q$-values of the BH procedure from observed $p$-values: \citep{storey2002direct, storey2003statistical}

\begin{enumerate}
\item Sort $\left(\breve{p}^{(1)}, \breve{p}^{(2)}, ..., \breve{p}^{(m)}\right)$ from the smallest to the largest value based on observed p-values $\left(\breve{p}_{1}, \breve{p}_{2}, ..., \breve{p}_{m}\right)$.
\item Set $q^{(m)} = \breve{p}^{(m)}$. 
\item Compute $q^{(i)} = \min \left[ \breve{p}^{(i)}\times m/i, q^{(i+1)} \right]$, for $1 \leq i<m$. 
\item Transfer $\left(q^{(1)}, q^{(2)}, ..., q^{(m)} \right)$ back to $\left(q_{1}, q_{2}, ..., q_{m} \right)$ based on the sorting index of $\left(\breve{p}_{1}, \breve{p}_{2}, ..., \breve{p}_{m}\right)$.
\end{enumerate}

In general, this control is also valid when those \textit{p}-values are positively correlated.\citep{benjamini2001control} The Benjamini-Yekutieli (BY) procedure is available to handle general forms of dependency.\citep{benjamini2001control} Based on the illustration in the previous Section \ref{sec:first}, we obtain decision functions by not necessarily $p$-values. In the next section, we introduce an algorithm to combine two layers, and to streamline the workflow.

\subsection{The whole story} 

In this section, we provide a complete picture of our SAFE framework based on the two layers discussed above. Algorithm \ref{alg_SAFE} streamlines the process of evaluating all SAs to check for potential safety concerns. Step 1 implements the procedure of the first layer in Section \ref{sec:first} to provide adjustment within each SA. Next, Step 2 follows the second layer in Section \ref{sec:second} to control FDR across all SAs. Finally, Step 3 conducts a decision-making process to evaluate if a specific SA has any safety findings. 

\begin{algorithm}
	\caption{Two-layer Synergy Area with FDR-controlled Evaluation (SAFE) Framework.}
	\label{alg_SAFE}
	\begin{tabbing}
		Step 1. For each $\text{SA}_i$, $i = 1, ..., m$, do\\
		\quad Step 1.1. Compute the raw $p$-value $p_{i, j}$ for each of the AE variable $j$ within this $\text{SA}_i$ ($j = 1, ..., n_i$). \\
		\quad Step 1.2. Compute the adjusted $p$-values $\left(\widetilde{p}_{i, 1}, \widetilde{p}_{i, 2}, ..., \widetilde{p}_{i, n_i}\right)$ from $n_i$ raw $p$-values based on the Holm procedure.\citep{holm1979simple}\\
		\quad Step 1.3. Sort adjusted $p$-values in ascending order, and assign the second smallest adjusted $p$-value to $\breve{p}_i$ as $\breve{p}_i = \widetilde{p}^{(2)}_{i}$. \\
		End \\
		Step 2. Compute $q$-values $\left({q}_{1}, {q}_{2}, ..., {q}_{m}\right)$ from $\left(\breve{p}_{1}, \breve{p}_{2}, ..., \breve{p}_{m}\right)$ from all $m$ SAs based on BH \citep{benjamini1995controlling} or BY \citep{benjamini2001control} in Section \ref{sec:second}. \\   
		Step 3. Declare that an $\text{SA}_i$ ($i = 1, ..., m$) has safety signals, if ${q}_{i} \leq \alpha$.  
	\end{tabbing}
\end{algorithm}

\section{Some alternatives}
\label{sec:alter}

An appealing feature of SAFE is that it robustly draws potential safety conclusions based on substantial evidence. The alternative hypothesis $G_1$ in (\ref{equ:g1_new}) is formulated as the intersection of at least two elementary alternative hypotheses. On the other hand, some previous works discussed in Section \ref{sec:intro} mainly utilize the alternative hypothesis $K_1$ in (\ref{equ:g1_ori}) if there is at least one elementary alternative hypothesis within all AEs or a cluster of AEs. \citep{berry2004accounting, dunson2008bayesian, berry2010bayesian, huque2010multiplicity, mehrotra2012flagging, dmitrienko2013traditional, norton2016perspective, menyhart2025multiplicity}

The specific choice between these two types of setups depends on the clinical context of a specific problem. While acknowledging a relatively fundamental difference in the underlying assumption, we consider some direct methods and compare their performances with SAFE in the case studies of Section \ref{sec:real}. Such direct methods focus on AE variables, and declare that a specific AE variable has significant findings if its multiplicity-adjusted $p$-value is smaller than or equal to the nominal level. The adjustment method has many flexible choices, such as the Holm procedure \citep{holm1979simple} to control FWER, or the BH procedure \citep{benjamini1995controlling} to control FDR. 

\section{Numerical studies}
\label{sec:sim}

In this section, we conduct some simulation studies to evaluate the FDR control of the proposed SAFE framework. We don't further evaluate the error rate control of direct methods in Section \ref{sec:alter}, because such properties have been well studied by previous works.\citep{holm1979simple, benjamini1995controlling, benjamini2001control} Comparative analyses are performed in Section \ref{sec:real}. 

The number of SAs $m$ is evaluated at $5$ or $10$, and the number of AE variables $n$ within $\text{SA}_j$ is $15$. The number of simulation iterations is $100,000$. The nominal level for FDR control is set as $\alpha = 5\%$. We simulate test statistics of all $m \times n$ AE variables from a multivariate normal distribution, with a standard deviation of $1$ for all variables. The correlation matrix has a compound symmetry structure, with $\rho_v$ as the correlation coefficient between AE variables within a specific SA, and $\rho_r$ for AE variables across different SAs. The mean of this multivariate normal distribution is $\left(\boldsymbol{\mu}_1, \boldsymbol{\mu}_2, ..., \boldsymbol{\mu}_m \right)$, where $\boldsymbol{\mu}_i$ ($i = 1, ..., m$) as a vector of length $n$ for means of $n$ AE variables within $\text{SA}_i$. To facilitate demonstration, we set $\boldsymbol{\mu}_i = \boldsymbol{\mu}_0 + w_i$, where $\boldsymbol{\mu}_0$ is a common mean vector of length $n$ for all SAs, and $w_i$ is an SA-specific scalar variable. Their specific values are specified later. The unadjusted $p$-value $p_{i, j}$ for $\text{SA}_i$ and AE variable $j$ is further computed as the upper tail of the cumulative distribution function of a standard normal distribution based on the corresponding test statistic. 

We first focus on the setup with $m=5$. On the choice of $\boldsymbol{\mu}_0$, there are two scenarios evaluated,
\begin{itemize}
	\item M1: $\boldsymbol{\mu}_0$ = $\boldsymbol{0}_{15}$,
	\item M2: $\boldsymbol{\mu}_0$ = (6, $\boldsymbol{0}_{14}$),
\end{itemize}
where $\boldsymbol{0}_{a}$ is a row vector of $0$ with $a$ elements. If $w_i=0$, then $\boldsymbol{\mu}_i = \boldsymbol{\mu}_0$ are under the null hypothesis $G_0^{(i)}$ for both M1 and M2 of $\boldsymbol{\mu}_0$. Specifically, all elementary null hypotheses in M1 are true, and only the first elementary alternative hypothesis in M2 is true. 

On the choice of $w_i$'s, we consider the following three settings:
\begin{itemize}
	\item S1: $w_1 = w_2 = w_3 = w_4 = w_5 = 0$,
	\item S2: $w_1 = 3$, $ w_2 = w_3 = w_4 = w_5 = 0$,
	\item S3: $w_1 = w_2 = 3$, $ w_3 = w_4 = w_5 = 0$ 
\end{itemize}
If $\boldsymbol{\mu}_0$ is under M1 or M2, then S1 means that none of the 5 SAs have true safety signals or findings. S2 has the first $\text{SA}_1$ with safety concerns, and S3 has $\text{SA}_1$ and $\text{SA}_2$ with signals. 

In Table \ref{tab:sim_1}, we evaluate 4 blocks of settings, with a combination of $\boldsymbol{\mu}_0$ and $w$ as: M1 + S1, M2 + S1, M2 + S2, and M2 + S3. Within each of the 4 blocks, we consider varying values of $\rho_r$ and $\rho_v$ to reflect positive dependency, independence, and negative dependency. Of note, $-1/74 = -1/(n \times m - 1)$ is the smallest value for the correlation coefficient to make a compound symmetric matrix positive definite. 

Table \ref{tab:sim_1} first reports probabilities of wrongly rejecting a null hypothesis $G_0^{(i)}$ when it is true, based on the first layer of SAFE in Section \ref{sec:first}. Therefore, such a probability is not displayed for $G_0^{(1)}$ under the third block with $\text{SA}_1$ under the alternative hypothesis, and is not displayed for $G_0^{(1)}$ and $G_0^{(2)}$ in the fourth block when both $\text{SA}_1$ and $\text{SA}_2$ are under alternative hypotheses. Next, FDR is presented based on the second layer of SAFE in Section \ref{sec:second}.

In the first block with M1 and S1, probabilities of wrongly rejecting individual $G_0^{(i)}$'s are below the nominal $\alpha = 5\%$. This is expected, because all AE variables have no safety concerns within a specific SA. The associated testing procedure for hypotheses in (\ref{equ:g0_new}) and (\ref{equ:g1_new}) can be conservative, as discussed in Section \ref{sec:second}. The FDR in the next layer is also well controlled below $5\%$. In the next three blocks with M2, those probabilities of wrong rejection are closer to the nominal level of $5\%$. This observation is consistent with Theorem 2 in Berger (1982)\citep{berger1982multiparameter}, which investigates several conditions for the error rate to be close to the nominal level. A specific condition is that some elementary null hypotheses in (\ref{equ:g0_new}) have high probabilities of being rejected, e.g., the first AE variable with a mean of $6$ in M2. The FDR based on the BH procedure\cite{benjamini1995controlling} is controlled at the nominal level. In some negative dependency structures, FDR based on the BH procedure may be slightly inflated. The BY procedure \citep{benjamini2001control} can be naturally utilized in the second layer of SAFE to control FDR under arbitrary dependency. When it comes to the third block with four true null hypotheses of SA, and the fourth block with three true null hypotheses of SA, the FDR becomes smaller. This phenomenon is consistent with some general properties of FDR. For instance, FDR is upper bounded by the proportion of true null hypotheses among all hypotheses times the nominal level $\alpha$ under independence.\citep{benjamini1995controlling} Section 2 of Supplementary Materials perform additional analyses with varying correlation structures to show robustness of error rate control. 

In this article, we focus on SAFE with at least two AEs within an SA to identify potential safety findings per Step 1.3 of Algorithm \ref{alg_SAFE}. A direct method can utilize the smallest $p$-value instead. Results in Section 1 of Supplementary Materials show that this direct method has inflated FDR under some scenarios, because its associated alternative hypothesis $K_1$ in (\ref{equ:g1_ori}) is less restricted than $G_1$ in (\ref{equ:g1_new}) of interest. The probabilities of true positive findings of the direct method are larger than SAFE as expected. There is a trade-off between a more robust evidence with positive findings in at least two AEs within an SA of SAFE and its associated decreased power of true positive findings.  

A more generalized setup SAFE($l$) has an integer $l$ ($l \geq 2$) denoting the minimum number of AEs. For instance, SAFE(3) requires positive findings in at least three AEs within an SA. Additional simulation studies are performed in Section 3 of Supplementary Materials to show that the error rate of false positive findings and power of true positive findings decrease as $l$ increases. Of note, SAFE($l$) with $l \geq 2$ can properly control error rates at the nominal level. The balance here is between a more robust evidence of lower false positive rates (with a larger $l$) and its associated decreased power of true positive findings. For a particular problem, operating characteristics can be evaluated to choose a proper value of $l$ under varying design parameters, for example, $n$ as the number of AEs within an SA. Moreover, one can also evaluate some adaptive rules to setup $l$, for example, a positive integer by rounding $20\% \times n$. 

Table \ref{tab:sim_2} shows results with $m=10$ SAs. Due to the relatively large number of SAs, we do not show probabilities for individual SAs, which are consistent with Table \ref{tab:sim_1}. Similarly, FDR is well controlled at the nominal level of $5\%$, and decreases as the number of true null hypotheses of SA decreases (Table \ref{tab:sim_2}).  

\begin{table}[ht]
	\centering
	\begin{tabular}{cccccccccc}
		\hline
		& & & & \multicolumn{5}{c}{Probability of wrongly rejecting} & \\
		\cmidrule(lr){5-9}
		$\boldsymbol{\mu}_0$ & $w$ & $\rho_r$ & $\rho_v$ & $G_0^{(1)}$ & $G_0^{(2)}$ & $G_0^{(3)}$ & $G_0^{(4)}$ & $G_0^{(5)}$ & FDR \\
		\hline
		M1 &  S1 & 0 & 0 & 0.1\% & 0.1\% & 0.1\% & 0.1\% & 0.1\% & $<$0.1\% \\
		&    & 0 & -1/74 & 0.1\% & 0.1\% & 0.1\% & 0.1\% & 0.1\% & $<$0.1\% \\
		&    & 0 & 0.7 & 1.0\% & 1.0\% & 1.1\% & 1.1\% & 1.0\% & 1.0\% \\
		&    & -1/74 & 0 & 0.1\% & 0.1\% & 0.1\% & 0.1\% & 0.1\% & $<$0.1\% \\
		&    & -1/74 & -1/74 & 0.1\% & 0.1\% & 0.1\% & 0.1\% & 0.1\% & $<$0.1\% \\
		&    & -1/74 & 0.7 & 1.1\% & 1.2\% & 1.0\% & 1.0\% & 1.1\% & 1.1\% \\
		&    & 0.7 & 0.7 & 1.0\% & 1.0\% & 1.0\% & 1.0\% & 1.0\% & 0.6\% \\
		\\
		M2 &  S1 & 0 & 0 & 4.8\% & 4.8\% & 4.9\% & 4.8\% & 5.0\% & 5.0\% \\
		&    & 0 & -1/74 & 5.0\% & 4.9\% & 4.9\% & 4.9\% & 4.9\% & 5.0\% \\
		&    & 0 & 0.7 & 2.5\% & 2.4\% & 2.4\% & 2.4\% & 2.4\% & 2.9\% \\
		&    & -1/74 & 0 & 4.9\% & 5.0\% & 4.8\% & 4.8\% & 5.0\% & 5.0\% \\
		&    & -1/74 & -1/74 & 4.9\% & 4.9\% & 4.9\% & 4.9\% & 5.0\% & 5.0\% \\
		&    & -1/74 & 0.7 & 2.5\% & 2.5\% & 2.4\% & 2.5\% & 2.4\% & 2.9\% \\
		&    & 0.7 & 0.7 & 2.4\% & 2.5\% & 2.5\% & 2.5\% & 2.5\% & 1.9\% \\
		\\
		M2 &  S2 & 0 & 0 & - & 4.9\% & 4.8\% & 5.0\% & 4.9\% & 4.0\% \\
		&    & 0 & -1/74 & - & 4.9\% & 4.9\% & 4.9\% & 4.8\% & 4.0\% \\
		&    & 0 & 0.7 & - & 2.5\% & 2.5\% & 2.5\% & 2.4\% & 2.2\% \\
		&    & -1/74 & 0 & - & 4.9\% & 4.9\% & 4.8\% & 4.9\% & 3.9\% \\
		&    & -1/74 & -1/74 & - & 4.9\% & 4.9\% & 4.9\% & 4.9\% & 3.9\% \\
		&    & -1/74 & 0.7 & - & 2.5\% & 2.5\% & 2.5\% & 2.5\% & 2.2\% \\
		&    & 0.7 & 0.7 & - & 2.4\% & 2.5\% & 2.5\% & 2.5\% & 1.7\% \\
		\\
		M2 &  S3 & 0 & 0 & - & - & 4.9\% & 4.9\% & 4.9\% & 3.0\% \\
		&    & 0 & -1/74 & - & - & 4.8\% & 5.0\% & 5.0\% & 3.0\% \\
		&    & 0 & 0.7 & - & - & 2.4\% & 2.5\% & 2.5\% & 1.6\% \\
		&    & -1/74 & 0 & - & - & 4.9\% & 4.9\% & 5.0\% & 3.0\% \\
		&    & -1/74 & -1/74 & - & - & 4.9\% & 4.8\% & 4.9\% & 2.9\% \\
		&    & -1/74 & 0.7 & - & - & 2.6\% & 2.5\% & 2.4\% & 1.6\% \\
		&    & 0.7 & 0.7 & - & - & 2.4\% & 2.5\% & 2.4\% & 1.3\% \\
		\hline
	\end{tabular}
	\caption{Probability of wrongly rejecting $G_0$ in (\ref{equ:g0_new}) for each SA in the first layer of SAFE (Section \ref{sec:first}), and FDR across all SAs in the second layer of SAFE (Section \ref{sec:second}), under $m=5$ SAs. }
	\label{tab:sim_1}
\end{table}

\begin{table}[ht]
	\centering
	\begin{tabular}{cccccccccc}
		\hline
		$\boldsymbol{\mu}_0$ & $w$ & $\rho_r$ & $\rho_v$ & FDR \\
		\hline
		M1 & S1 & 0 & 0 & $<$0.1\% \\
		&  & 0 & -1/149 & $<$0.1\% \\
		&  & 0 & 0.7 & 0.9\% \\
		&  & -1/149 & 0 & $<$0.1\% \\
		&  & -1/149 & -1/149 & $<$0.1\% \\
		&  & -1/149 & 0.7 & 1.0\% \\
		&  & 0.7 & 0.7 & 0.6\% \\
		\\
		M2 & S1 & 0 & 0 & 5.0\% \\
		&  & 0 & -1/149 & 4.8\% \\
		&  & 0 & 0.7 & 3.1\% \\
		&  & -1/149 & 0 & 4.9\% \\
		&  & -1/149 & -1/149 & 5.0\% \\
		&  & -1/149 & 0.7 & 3.1\% \\
		&  & 0.7 & 0.7 & 1.6\% \\
		\\
		M2 & S2 & 0 & 0 & 4.4\% \\
		&  & 0 & -1/149 & 4.6\% \\
		&  & 0 & 0.7 & 2.7\% \\
		&  & -1/149 & 0 & 4.5\% \\
		&  & -1/149 & -1/149 & 4.4\% \\
		&  & -1/149 & 0.7 & 2.7\% \\
		&  & 0.7 & 0.7 & 1.7\% \\
		\\
		M2 & S3 & 0 & 0 & 4.0\% \\
		&  & 0 & -1/149 & 4.0\% \\
		&  & 0 & 0.7 & 2.4\% \\
		&  & -1/149 & 0 & 4.0\% \\
		&  & -1/149 & -1/149 & 3.9\% \\
		&  & -1/149 & 0.7 & 2.4\% \\
		&  & 0.7 & 0.7 & 1.6\% \\
		\hline
	\end{tabular}
	\caption{FDR across all SAs in the second layer of SAFE (Section \ref{sec:second}), under $m=10$ SAs. }
	\label{tab:sim_2}
\end{table}

\section{Case studies}
\label{sec:real}

In this section, we conduct two case studies based on real data from the Historical Trial Data (HTD) Sharing Initiative of TransCelerate BioPharma. We do not present patient-level information and do not disclose the sponsor's name. 

\subsection{Case study 1}

In the first case study, we focus on two medical conditions of Hidradenitis Suppurativa (HS) and Psoriasis (PsO). HS is a chronic, disabling, inflammatory skin disease characterized by recurrent, painful nodules, abscesses, draining fistulas, and scarring \citep{sabat2020hidradenitis, ackerman2025improvements}, while PsO is a chronic immune-mediated disease, and its most common variant plaque psoriasis is characterized by erythematous scaly patches or plaques that occur commonly on extensor surfaces.\citep{armstrong2020pathophysiology, strober2020efficacy}

The objective of this case study is to evaluate whether or not the safety profile of patients with HS is different from that of PsO. For the HS group, we include safety data from a clinical trial with patients who were diagnosed with HS and were randomized to receive placebo. For the PsO group, safety data from placebo patients diagnosed with PsO from another clinical trial are included. We implement our proposed SAFE framework to compare the safety of those two groups of data. 

The Synergy Area (SA) is chosen as the System Organ Class (SOC). There are $m=23$ SAs, and a total of $259$ AE variables. We obtain the raw $p$-value of a specific AE variable from the two-group proportional test without continuity correction based on the incidence rates of this AE from two groups. 

\begin{figure}
	\centering
	\begin{subfigure}[b]{0.9\textwidth}
		\centering
		\includegraphics[width=1\linewidth]{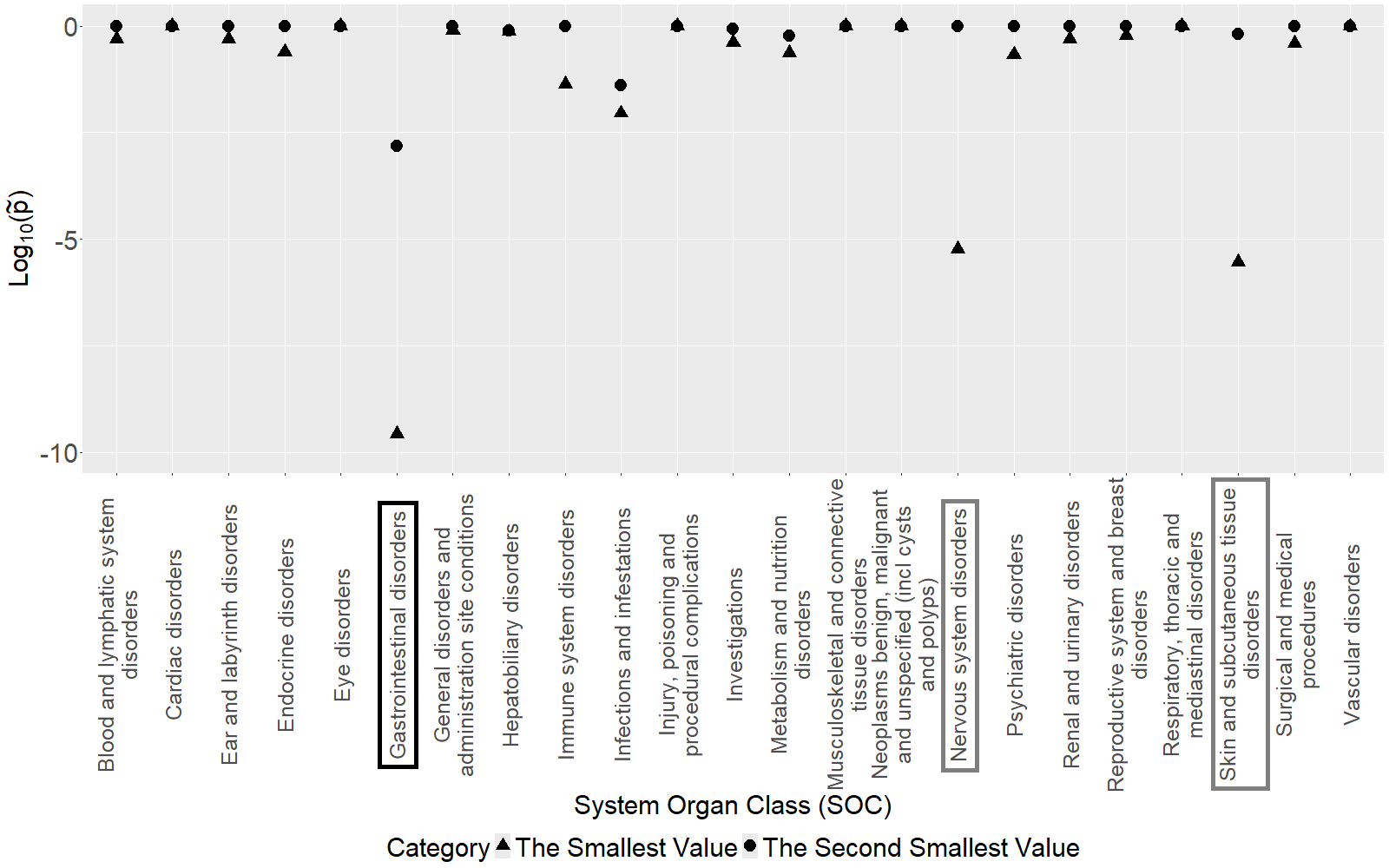}
		\caption{Adjusted $p$-values $\widetilde{p}$ from the first layer of SAFE.}
		\label{fig:case_1_p}
	\end{subfigure}%
	\\
	\bigskip
	\bigskip
	\bigskip
	\begin{subfigure}[b]{0.9\textwidth}
		\centering
		\includegraphics[width=1\linewidth]{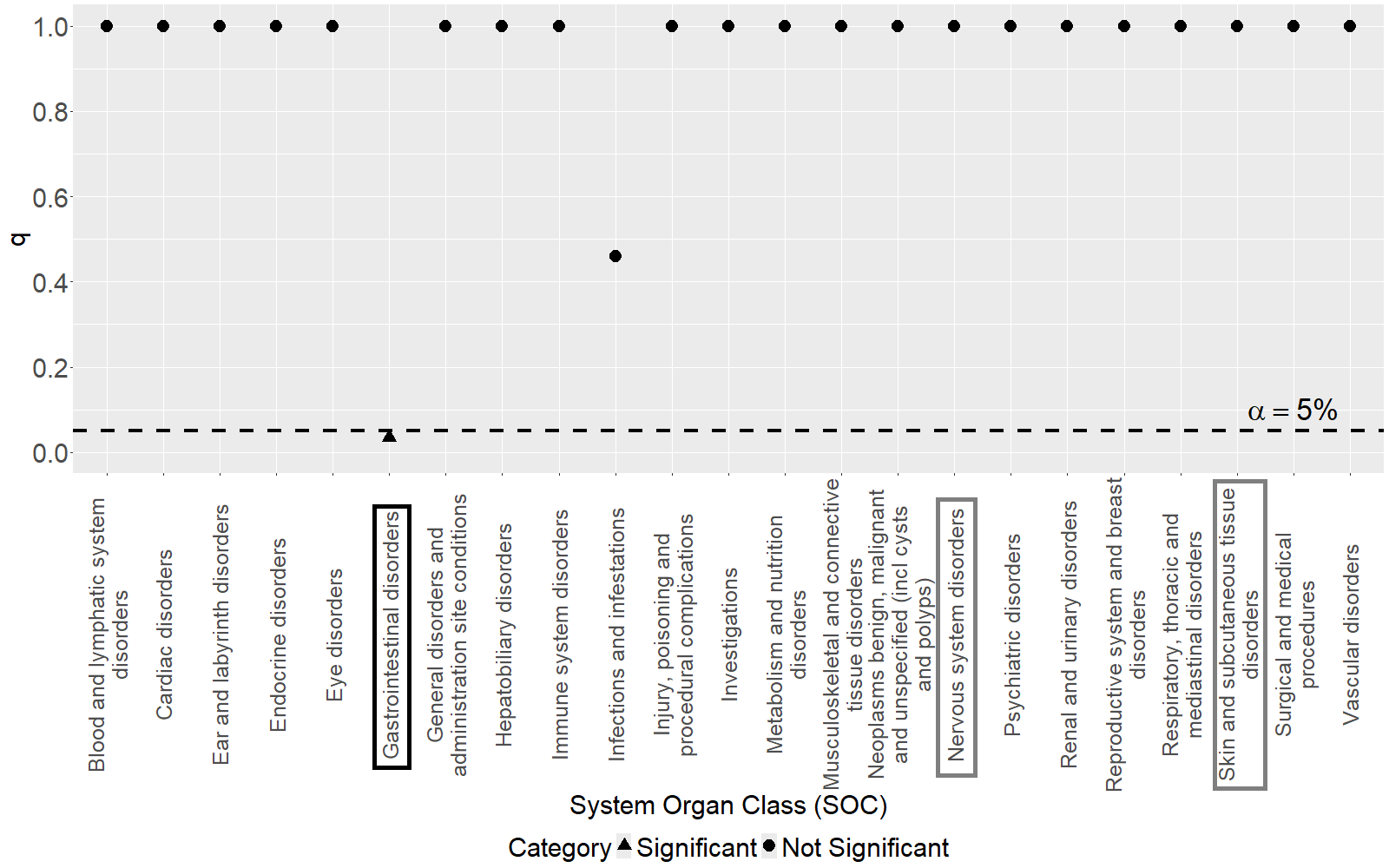}
		\caption{$Q$-values from the second layer of SAFE. }
		\label{fig:case_1_q}
	\end{subfigure}
	\caption{Case study 1 of comparing the safety profile of patients with Hidradenitis Suppurativa (HS) versus patients with Psoriasis (PsO). }
	\label{fig:case_1}
\end{figure}

In the first layer of SAFE (Section \ref{sec:first}), we calculate Holm-adjusted $p$-values $\left(\widetilde{p}_{i, 1}, \widetilde{p}_{i, 2}, ..., \widetilde{p}_{i, n_i}\right)$ within a specific $\text{SA}_i$ ($i$th SOC, $i = 1, ..., 23$) based on raw $p$-values. Figure \ref{fig:case_1_p} shows the smallest adjusted $p$-value $\widetilde{p}_i^{(1)}$ in triangle, and the second smallest value $\widetilde{p}_i^{(2)}$ in circle, for each of the $23$ SAs. The adjusted $p$-values on the y-axis are transformed by a logarithm with a base of 10. The SOC of ``gastrointestinal disorders'' with a black rectangle outline has the smallest $\widetilde{p}_i^{(2)}$ ($\text{log}_{10}\left[\widetilde{p}_i^{(2)}\right] = -2.81$) among all SAs (Figure \ref{fig:case_1_p}). Two other SAs with gray rectangle outlines, ``nervous system disorders'' and ``skin and subcutaneous tissue disorders'' have small $\widetilde{p}_i^{(1)}$'s with $\text{log}_{10}\left[\widetilde{p}_i^{(1)}\right]$ at $-5.23$ and $-5.53$, respectively, but their $\widetilde{p}_i^{(2)}$'s are close to $0$ (Figure \ref{fig:case_1_p}). 

In the second layer of SAFE (Section \ref{sec:second}), we compute $q$-values of all SAs $\left(\widetilde{p}_1^{(2)}, \widetilde{p}_2^{(2)}, ..., \widetilde{p}_m^{(2)} \right)$ based on the BH procedure. Figure \ref{fig:case_1_q} shows that the SA of ``gastrointestinal disorders'' with a black rectangle outline has a $q$-value of $3.6\%$, which is smaller than the nominal level $\alpha = 5\%$. This finding suggests that the safety profile of the HS group is significantly different from the PsO group in this SA. Motivated by this finding, one can conduct additional medical review to evaluate if a particular group has worse safety. Additionally, those two SAs in gray outlines with even smaller $\widetilde{p}_i^{(1)}$ do not have significant safety findings identified by our SAFE framework, because there are no additional substantial evidence (i.e., sufficiently small $\widetilde{p}_i^{(2)}$) to confirm potential safety signals. 

For those alternative methods discussed in Section \ref{sec:alter}, they directly apply multiplicity adjustment to raw $p$-values from all AE variables. Specifically, the direct Holm procedure identifies two AEs with significant safety findings within the ``gastrointestinal disorders'' in a black outline, and one significant AE within each of the two SAs with gray outlines and an additional SA of ``infections and infestations''. The direct BH approach has exactly the same finding as the direct Holm approach, except that the direct BH finds two additional significant AEs within the SA of ``infections and infestations''. This is expected given a generally larger power of the BH procedure with FDR control than the Holm procedure with FWER control.\citep{benjamini1995controlling}

SAFE finds that safety profile of gastrointestinal disorders in HS patients from one trial is different from PsO patients from the other trial. Some previous studies showed that HS patients might be more likely to experience gastrointestinal dysfunction as compared to a general population or patients with other skin conditions. \citep{chen2019association, cartron2019comorbidities, chang2026risk} Such results are not confirming evidence, but serve as exploratory findings to motivate more research in this area to generate more substantial evidence. 

As a summary, our proposed SAFE framework identifies potential safety findings in one SA of ``gastrointestinal disorders'' with relatively substantial evidence, while alternative direct methods find two additional SAs based on relatively extreme data. Such results showcase a major advantage of the SAFE framework to robustly characterize the safety profile with dual evidence within an SA. 

\subsection{Case study 2}

In the second case study, we consider two clinical trials in Atopic Dermatitis (AD), which is a chronic inflammatory skin disease that is characterized by intense itching and recurrent eczematous lesions.\citep{weidinger2016atopic} SAFE and some direct methods are implemented to identify potential safety signals. While heterogeneity may exist for patients in two trials, it is likely that there is no significant difference in safety profiles between them with the same indication. 

\begin{figure}
	\centering
	\begin{subfigure}[b]{0.9\textwidth}
		\centering
		\includegraphics[width=1\linewidth]{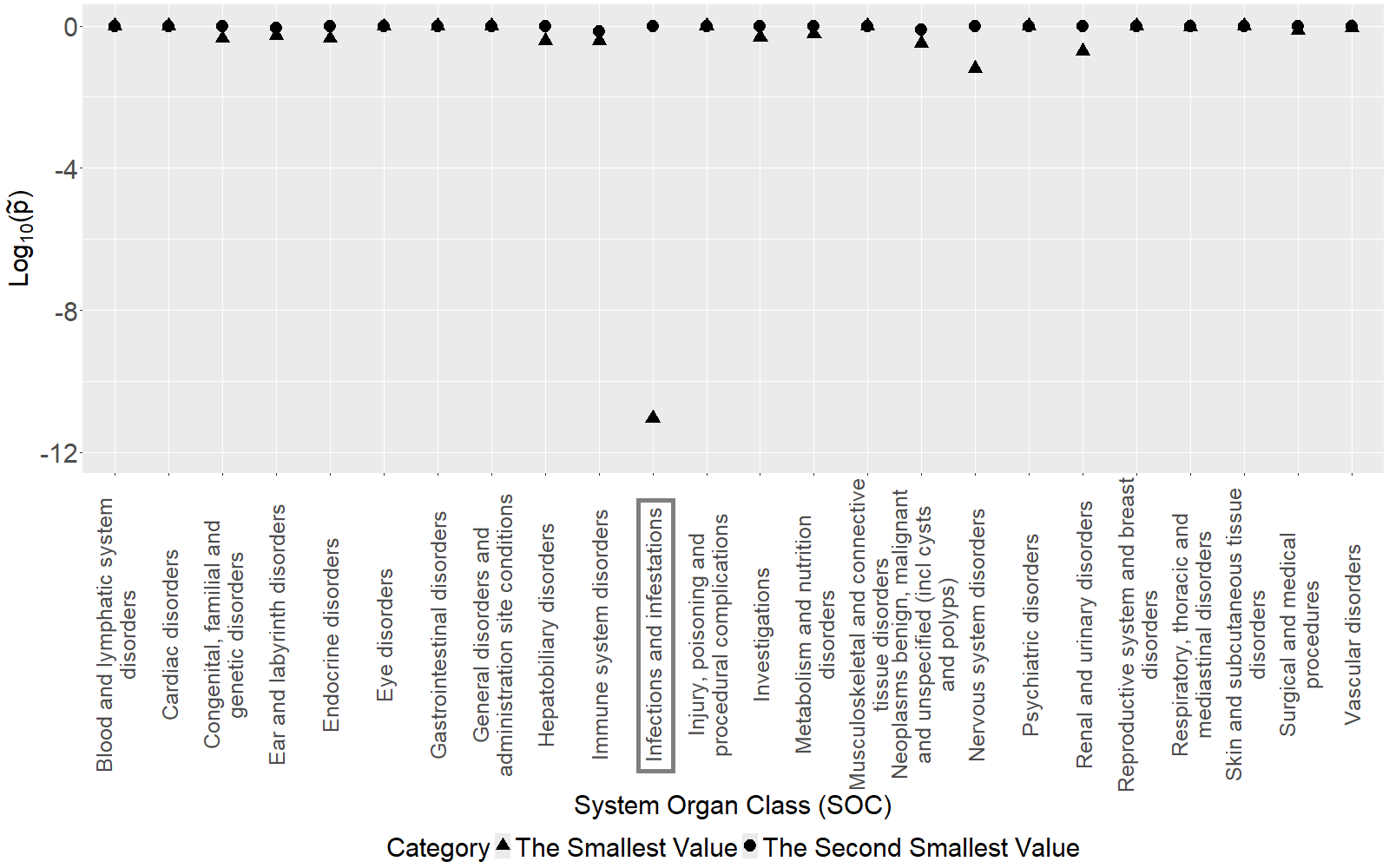}
		\caption{Adjusted $p$-values $\widetilde{p}$ from the first layer of SAFE.}
		\label{fig:case_2_p}
	\end{subfigure}%
	\\
	\bigskip
	\bigskip
	\bigskip
	\begin{subfigure}[b]{0.9\textwidth}
		\centering
		\includegraphics[width=1\linewidth]{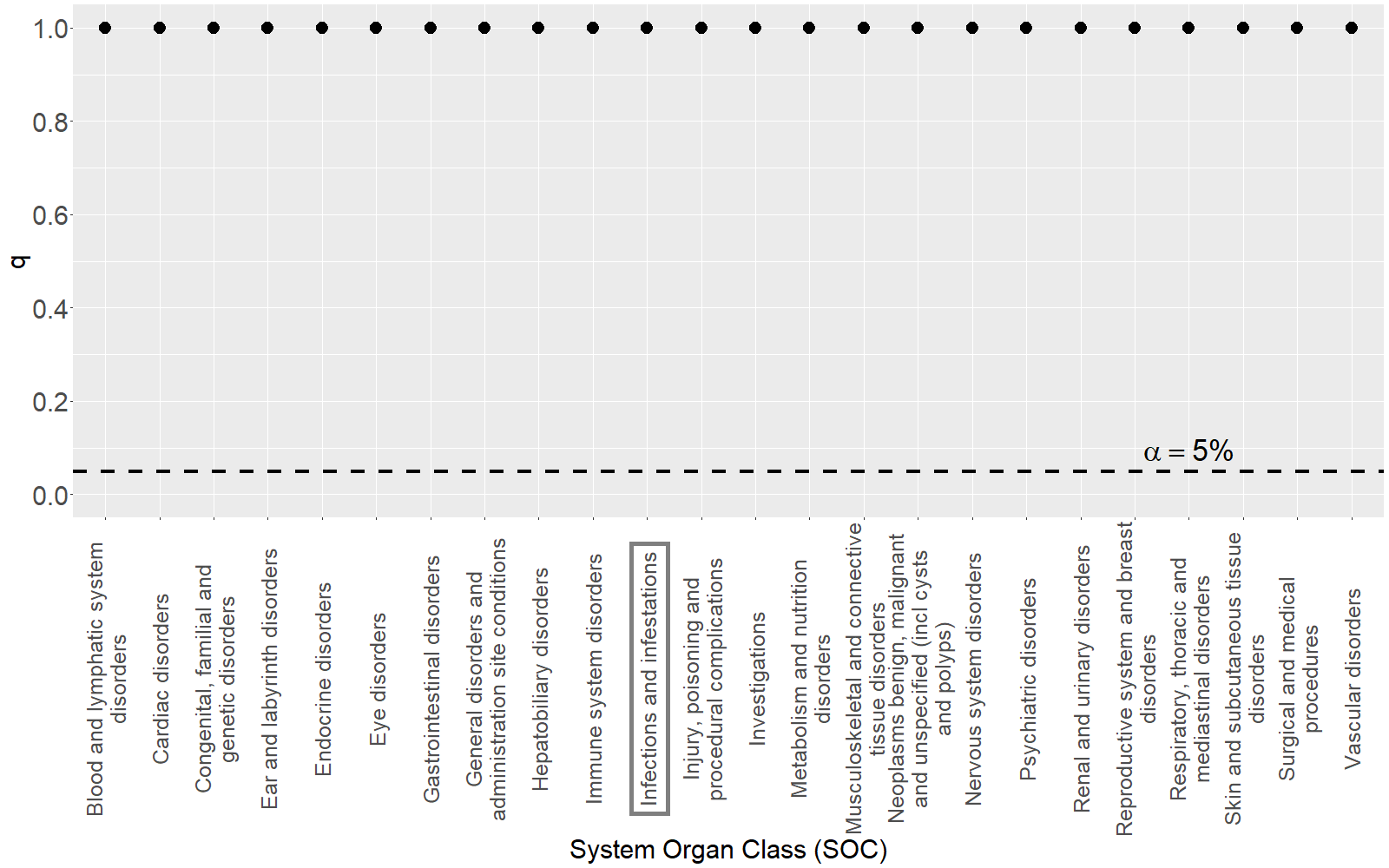}
		\caption{$Q$-values from the second layer of SAFE.}
		\label{fig:case_2_q}
	\end{subfigure}
	\caption{Case study 2 of comparing the safety profile of patients receiving placebo from two clinical trials in treating Hidradenitis Suppurativa (HS).}
	\label{fig:case_2}
\end{figure}

Figure \ref{fig:case_2_p} presents the adjusted $p$-values $\widetilde{p}_i$ from the first layer of SAFE. All SAs have relatively large second smallest adjusted $p$-values $\widetilde{p}_i^{(2)}$ in black dots. The smallest adjusted $p$-values $\widetilde{p}_i^{(1)}$ in black triangles of one SA (labeled by gray outline) is relatively extreme with $\text{log}_{10}\left[\widetilde{p}_i^{(1)}\right]$ at $-11.05$. When it comes to the second layer of SAFE in Figure \ref{fig:case_2_q}, $q$-values from all SAs are larger than the nominal level of $5\%$, indicating that there are no significant differences in safety profile between these two groups. On the other hand, both the direct Holm and the direct BH method identify the same AE with that SA (labeled by a gray outline) with significant findings.

As a summary, in this case study where two groups tend to have similar safety profiles, our proposed SAFE framework does not identify substantial evidence to claim that there are significant signals, while some direct methods have findings based on some relatively extreme data. 

\section{Discussion}
\label{sec:disc}

In this article, we propose a Synergy Area with FDR-controlled Evaluation (SAFE) structural framework to evaluate the safety profile in clinical trials with three major appealing features: a controlled error rate, integration of clinical or domain knowledge, and robust conclusions based on substantial evidence. However, potential findings from SAFE are not the final conclusions, and should be confirmed by additional clerks' review. It provides quantitative and automatic assistance to the traditional large-scale manual review to ensure prioritization, increase efficiency, and also trigger sparks to new scientific problems for further investigations. In this article, we mainly focus on the evaluation of safety profiles from completed clinical trials. There are many other potential applications: safety monitoring of ongoing trials, drug safety surveillance of post-marketing, and assistance to regulatory review of safety data.

The proposed SAFE framework relies on substantial and robust evidence to evaluate safety profile, and therefore also requires high-quality data. To address a potential issue of missing data, a future work of SAFE is to incorporate some missing data handling strategies. Moreover, data monitoring is also critical to ensure that the input safety data are accurate. 

The introduced SAFE framework is general and flexible to accommodate the context of a specific problem. First of all, the nominal level $\alpha$ to control FDR in the second layer across all SAs (Section \ref{sec:second}) reflects the tolerance level on the false positive error. It can be set to a relatively large value if the study team wants to identify potential safety signals more aggressively while acknowledging an increased error rate. Secondly, in the current first layer within an SA (Section \ref{sec:first}), one needs at least two elementary findings to provide substantial evidence. This can readily be generalized to a setting with at least four elementary findings to provide a more solid support. Additionally, more layers can be added to establish a multi-layer SAFE. For instance, a three-layer structure is formulated based on terminologies in MedDRA (Medical Dictionary for Regulatory Activities)\cite{meddra}, where AEs are first grouped to ``Lowest Level Terms'' (LLTs) based on clinical knowledge, and then LLTs are clustered to ``High Level Terms'' (HLTs). 

There are also some extensions based on the proposed SAFE framework in this article. On the one hand, SAFE is able to act as a platform to incorporate other parameters of interest, for example, a certain composite score to evaluate both safety and efficacy in the benefit-risk assessment.\citep{he2016benefit} On the other hand, SAFE can serve as a building block to other infrastructures. For instance, in an aggregated safety evaluation tool, SAFE and its variants focus on different aspects or dimensions of a safety event to provide distilled conclusions. The proposed SAFE is applied to two clinical studies in this work, but it can be generally implemented under other settings, e.g., postmarketing surveillance. 

\section*{Acknowledgments}

The authors thank the editor, AE, and two reviewers for their constructive comments that significantly improved this article. This manuscript was supported by AbbVie Inc. AbbVie participated in the review and approval of the content. All authors are employed by AbbVie Inc., and may own AbbVie stock. 

\section*{Supplementary Materials}

The R code to replicate results in Section \ref{sec:sim} is available at \url{https://github.com/tian-yu-zhan/SAFE}. 

\section*{Data Availability}

Data in Section \ref{sec:real} were obtained through TransCelerate Biopharma Inc.’s Historical Trial Data (HTD) Sharing Initiative, hosted on TransCelerate’s DataCelerate platform. Contributing companies include: AbbVie, Inc., Amgen, Inc., Astellas Pharma Global Development, Inc., AstraZeneca AB, Boehringer Ingelheim International GmbH, Bristol Myers Squibb Company, EMD Serono Research \& Development Institute, Inc., Genentech, Inc., GlaxoSmithKline LLC, Eli Lilly and Co., Janssen Research \& Development, LLC, Novartis Pharma AG, Novo Nordisk A/S, Pfizer Inc., Sanofi USA, Shionogi \& Co., Ltd., and UCB Biosciences GmbH.

\bibliographystyle{Chicago}
\bibliography{SAFE_ref.bib}

%\section*{Author Biography}
%
%\begin{biography}{\includegraphics[width=66pt,height=86pt,draft]{empty}}{\textbf{Author Name.} This is sample author biography text this is sample author biography text this is sample author biography text this is sample author biography text this is sample author biography text this is sample author biography text this is sample author biography text this is sample author biography text this is sample author biography text this is sample author biography text this is sample author biography text this is sample author biography text this is sample author biography text this is sample author biography text this is sample author biography text this is sample author biography text this is sample author biography text this is sample author biography text this is sample author biography text this is sample author biography text this is sample author biography text.}
%\end{biography}

\end{document}